\documentstyle[12pt]{article}
\newcommand{\beq}{\begin{equation}}
\newcommand{\eeq}{\end{equation}}
\newcommand {\be}{\begin{eqnarray}}
\newcommand {\ee}{\end{eqnarray}}
\begin{document}
\setcounter{footnote}{0}
\renewcommand{\thefootnote}{\fnsymbol{footnote}}

\rightline{RUB-TP2-05/01}
\rightline{}
\vspace{1cm}
\begin{center}
{\bf\Large A note on inequalities for the masses of the lightest
$\pi\pi$ resonances in large $N_c$ QCD  } \\ \vspace{1cm}
\setcounter{footnote}{0}
\renewcommand{\thefootnote}{\arabic{footnote}}

{\bf\large
M.V.~Polyakov{\footnote{e-mail:maximp@tp2.ruhr-uni-bochum.de}} }\\
{\it Petersburg Nuclear Physics Institute, Gatchina,\\
 18835 St.Petersburg, Russia} \\[.1cm]
{\it Institute for Theoretical Physics II \\
 Ruhr University, 44780 Bochum, Germany} \\[.1cm]
{\bf\large
V.~V.~Vereshagin{\footnote{e-mail:vvv@av2467.spb.edu}}  }\\
{\it Institute of Physics\\
 St.~Petersburg State University \\
 St.~Petersburg, 198504, Russia}
\end{center}

\begin{abstract}
We derive and analyse inequalities relating masses of the lightest
$\pi\pi$ resonances ($\rho$ and $\sigma$) to low energy couplings
of the effective chiral Lagrangian in the limit of large number of
colours.
\end{abstract}

%\newpage
%\noindent
%\section{Introduction}
%\setcounter{equation}{0}
%\renewcommand{\thefootnote}{\arabic{footnote}}

\noindent
{\bf 1.} The issue of the existence and interpretation
of the light scalar resonance
(we call it as $\sigma$ in what follows)
is one of the most controversial questions in the meson spectroscopy
(for a review of the scalar meson spectroscopy
see the note on scalar mesons by S.~Spanier and N.~Tornqvist in
Review of Particle Physics
\cite{Spanier:1998jm}\footnote{
Related problems are also discussed in the review article
\cite{VMP:2000}.}.
Far from complete list of experimental and theoretical
papers devoted to the
$\sigma$--meson
\cite{Tornqvist:1995}-\cite{Kusaka:1993yx}
(see also
\cite{VMP:2000}
and references therein)
shows that this topic attracts considerable interest.

In these notes we analyse the sum rules relating low energy
constants (LECs) of the effective chiral lagrangian (EChL)
to the resonance spectrum parameters in the limit of large
number of colours
\cite{BolPolVer,BolManPolVer}.
\footnote{In
\cite{AVVVVV}
it is shown that those sum rules can be derived from
the general postulates of the effective theory without
referring to large-$N_c$ limit; see also
\cite{Zuoz99}.}
We shall show that from these sum rules one can derive a set
of inequalities, e.g. such as:
\be
\label{ner-main}
M_\sigma^2(3 L_2+L_3)+M_\rho^2 L_2
\leq \frac{F_\pi^2 }{4} \,,
\ee
where $L_i$ are the coupling constants of the
fourth order EChL
\cite{GLsu3},
$M_\rho$ is the mass of the lightest isovector resonance
($\rho$ meson), $M_\sigma$
is the mass of the lightest isoscalar resonance
($\sigma$ meson), and
$F_\pi\approx 93$~MeV
is the pion decay constant. This inequality, apart from
applications for estimates of the
$\sigma$-meson
mass from above, demonstrates that properties of the resonance
spectrum are in close relations with properties of chiral symmetry
breaking. Below we give derivation of the inequality
(\ref{ner-main})
as well as its enhancements.

%\subsection{Large $N_c$ sum rules}
\vspace{0.2cm}
\noindent
{\bf 2.}
In the ref.
\cite{BolPolVer,BolManPolVer}
the following set of the large
$N_c$
sum rules relating the constants of the effective chiral Lagrangian
$L_i$\footnote{Note that LECs $L_i$ are scale
independent in the large $N_c$ limit}
to the parameters of resonance spectrum has been derived:
\be
\nonumber
1+O(m_\pi^4)&=&
\sum \frac{F_0^2 V_0}{\left[M_0^2-2 m_\pi^2\right]^2} +
\sum \frac{F_0^2 V_1}{\left[M_1^2-2 m_\pi^2\right]^2}
\, ,\\
\nonumber
3L_2+L_3+\alpha m_\pi^2+O(m_\pi^4)&=& \frac{F_0^4}{4}
\sum \frac{V_0}{\left[M_0^2-2 m_\pi^2\right]^3}
\, ,\\
L_2 +\beta m_\pi^2+O(m_\pi^4)&=&
\frac{F_0^4}{4}
\sum \frac{V_1}{\left[M_1^2-2 m_\pi^2\right]^3}
\, .
\label{sr}
\ee
Here
$M_I$
are the masses of pion-pion resonances with isospin
$I$, and $V_I$ ---
the corresponding residues. The latter are related
to the
$\pi\pi$
resonance width
$\Gamma(R\rightarrow \pi\pi)$
via:
\be
V_0&=&\frac23 16\pi (2J+1) \frac{M_0^2}{\sqrt{M_0^2-4 m_\pi^2}}
\Gamma(R\rightarrow \pi\pi)\, ,
\label{V0}
\\
V_1&=& 16\pi (2J+1) \frac{M_1^2}{\sqrt{M_1^2-4 m_\pi^2}}
\Gamma(R\rightarrow \pi\pi)\, ,
\ee
where $J$ is the resonance spin. The constant
$F_0\approx 88$~MeV
is the pion decay constant in the chiral limit. The constants
$\alpha,\beta$
are related to low energy coefficients (LECs) of the sixth order
EChL. We use estimates for the LECs of the sixth order EChL
obtained in
ref.~\cite{PolVer,Polyakov:1996hv}
from the chiral expansion of the dual (string) models:
\be
\alpha m_\pi^2 \approx 0.18\cdot 10^{-3},\quad
\beta m_\pi^2\approx 0.05 \cdot 10^{-3}\, .
\ee
{}From the sum rules eqs.
(\ref{sr})
one can immediately obtain the following obvious inequalities:
\be
\nonumber
&&3L_3+L_2 +\alpha m_\pi^2> 0 \, , \\
&&L_2 +\beta m_\pi^2> \frac{V_\rho F_0^4}{4(M_\rho^2 -2 m_\pi^2)^3}
\approx 1.66\cdot
10^{-3}\,,
\label{ner1}
\ee
where $M_\rho$ and $V_\rho$
stand for the mass and residue of the lightest isovector
$\pi\pi$ resonance
($\rho$--meson).
Further noting that
\beq
\sum_{I=0} \frac{V_0}{\left[M_0^2-2m_\pi^2\right]^k} >
\left[M_\sigma^2-2 m_\pi^2\right] \sum_{I=0}
\frac{V_0}{\left[M_0^2-2 m_\pi^2\right]^{k+1}} \, ,
\eeq
where
$k \geq 2$ and $M_\sigma$
is a mass of the lightest isoscalar (scalar) resonance,
we obtain the following inequality:
\be
\label{ner2}
&&M_\sigma^2\left(3 L_3+L_2+\alpha m_\pi^2\right)+
M_\rho^2 \left(L_2+\beta m_\pi^2\right)
< \frac{F_0^2 }{4}+2 m_\pi^2\left(4L_2+L_3\right) \, .
%
%&&V_\sigma <
%\frac{M_\sigma^4}{F_0^2}\Bigl( 1-
%\frac{4M_\rho^2}{F_0^2} L_2 \Bigr) \,.
%\label{ner3}
\ee
This inequality provides us with a nice example of nontrivial
relations between the parameters of resonance spectrum and low
energy constants of EChL. The model independent
large-$N_c$ inequality
(\ref{ner2})
can be used
%apart from practical applications
for the estimates of the
$\sigma$--meson
mass from above (see below),
as well as
%can be used
for consistency checks
of various models of low--energy QCD in the large-$N_c$ limit.

Parameters of the EChL in the large-$N_c$ limit have been
calculated in various models of the low--energy QCD
\cite{DiaEid,Bal,And,DiaPet,Espriu,Bij}.
We shall use parameters from the analysis of the
EChL coupling constants in the large-$N_c$ limit done in
\cite{Bij}
(the error bars take into account different values of the
constants obtained in the fits performed in
\cite{Bij}):
\be
\nonumber
L_2&=&~(1.6\pm 0.1)\cdot 10^{-3} \, ,\\
L_3&=&-(4\pm 1)\cdot 10^{-3} \, ,
\label{nc:parameters}
\ee
These values are close to those obtained from the phenomenological
analysis
\cite{GLsu3,BijColGas},
what shows that the
$1/N_c$
corrections to low energy coefficients $L_i$ are rather small.

Due to the inequality
eq.~(\ref{ner1})
the value of
$L_2$ can not be below
$1.63\cdot 10^{-3}$,
therefore we shall use this minimal value of
$L_2=1.63\cdot 10^{-3}$
lying in the range given by the
eq.~(\ref{nc:parameters})
\footnote{Note that for larger values of
$L_2$
the bounds on
$M_\sigma$
discussed below are stronger}.
The error of calculation of $L_3$ is bigger.
Also, the errors of $L_2$ and $L_3$ are strongly correlated.
In order to make an estimations of the $M_\sigma$ based
on inequality
eqs.~(\ref{ner2})
we use first the relation
$2L_2+L_3=0$
which follows from integration of the non--topological
chiral anomaly
\cite{DiaEid,Bal,And}
and from the low--energy limit of the dual--resonance (string) models
\cite{PolVer}.
Using the above values of
$L_2$ and $L_3$
we obtain from
eqs.~(\ref{ner2}):
\be
M_\sigma &<& 770\ \mbox{\rm MeV},\quad {\rm if\ } 2L_2+L_3=0\, .
\ee
This is the upper bound for the lightest isoscalar resonance
if one assumes the relation
$2L_2+L_3=0$.
To consider the more general case, we derive the upper limit on
$M_\sigma$
as a function of the parameter
$\Delta$
defined as follows:
\be
\Delta=-\frac{2L_2+L_3}{L_2}\, .
\label{deltadef}
\ee
The value of this parameter is zero for EChL obtained by
integration of non-topological chiral anomaly
\cite{DiaEid,Bal,And,DiaPet}
as well as for EChL obtained by chiral expansion of the
dual--resonance (string) models
\cite{PolVer}.
In the large-$N_c$ based model of
ref.~\cite{Espriu}
the value of
$\Delta$
is fixed in terms of gluon condensate and constituent quark mass
$m_Q\approx 0.35$~GeV as
$
\Delta=\frac{\pi^2 \langle\frac{\alpha_s}{\pi} G^2\rangle}{5N_c m_Q^4}
\approx 0.3 \, .
$
The value of LECs obtained in
ref.~\cite{Peris} corresponds to
$\Delta=5/8=0.625$.
In any case the value of
$\Delta$
can not exceed unity due to the inequality
(\ref{ner1}).
Experimentally, the parameter
$\Delta$ is constrained by the ratio
of the D-wave pion scattering lengths:
\be
\Delta = -3\ \frac{a_2^2}{a_2^0}+O(m_\pi^2) \approx -0.2 \pm 0.6 \, ,
\ee
where we took the experimental values of the D-wave
scattering lengths
ref.~\cite{exppipi}.

Now it is easy to derive from the inequality
(\ref{ner2})
the upper bound for the
$\sigma$ meson mass
as a function of the parameter
$\Delta$. This function at small
values of
$\Delta$ takes the form
\be
\label{ner2num}
M_\sigma < 770 \left[1
 +0.42 \Delta + 0.29 \Delta^2+ O(\Delta^3)\right]~{\rm MeV}\, .
\ee
We see that the upper bound for the $\sigma$-meson mass is sensitive
to the sign of the parameter $\Delta$ (see definition (\ref{deltadef})).
Therefore the values of LECs of the fourth order EChL can give us a
valuable information about the lightest scalar meson in the spectrum
of QCD.

\vspace{0.2cm}
\noindent
{\bf 3.} In the case when one posseses an additional information
(masses and widths of resonances)
on the excited meson spectrum
(mesons heavier than $\sigma$ in the isoscalar
channel
and $\rho$ in isovector one)
the inequality (\ref{ner2}) can be enhanced. Let us call the
excited resonances for which we have additional information about
their masses and widths as {\it known}. With this additional information
the inequality (\ref{ner2}) can be enhanced as follows:
\be
\nonumber
&&M_\sigma^2\left(3 L_3+L_2+\alpha m_\pi^2
-\sum_{\rm known} \frac{F_0^4 V_0}{4\left[M_0^2-2m_\pi^2\right]^3}\right)+
M_\rho^2 \left(L_2+\beta m_\pi^2
-\sum_{\rm known} \frac{F_0^4
V_1}{4\left[M_1^2-2m_\pi^2\right]^3}\right)\\
\label{ner2e}
&&< \frac{F_0^2 }{4}
-\sum_{\rm known} \frac{F_0^4 V_0}{4\left[M_0^2-2m_\pi^2\right]^2}
-\sum_{\rm known} \frac{F_0^4 V_1}{4\left[M_1^2-2m_\pi^2\right]^2}\\
&&+2 m_\pi^2\left(4L_2+L_3
-\sum_{\rm known} \frac{F_0^4 V_0}{2\left[M_0^2-2m_\pi^2\right]^3}
-\sum_{\rm known} \frac{F_0^4 V_1}{2\left[M_1^2-2m_\pi^2\right]^3}
\right) \, .
\nonumber
\ee
For the numerical estimates we take as the {\it known} resonances
$f_2(1275)$ in the isoscalar channel and $\rho_3(1690)$ in the
isovector channel. We do not include other scalar and vector mesons
as their nature is not well established and it is not clear whether
their dynamics is ``leading" in large $N_c$. Taking the masses and
$\pi\pi$ widths of $f_2(1275)$ and $\rho_3(1690)$ from
\cite{Spanier:1998jm} we obtain the enhancement of the inequality
(\ref{ner2num})
\be
\label{ner2enum}
M_\sigma < 665 \left[1
 +0.44 \Delta + 0.33 \Delta^2+ O(\Delta^3)\right]~{\rm MeV}\, .
\ee
Obviously the inclusion of other resonances, e.g. $f_0(980), f_0(1370)$,
$f_0(1500)$, $\rho'$, $f_4$, etc. would lead to lower bound on the  mass
of
$\sigma$--meson.

\vspace{0.2cm}
\noindent
{\bf 4.} To summarize, we derive the inequalities for the masses
of the lightest $\pi\pi$ resonances in the limit of large number of
colours ($N_c\to\infty$), see eqs.~(\ref{ner2},\ref{ner2e}).
These inequalities put an upper bound on the mass of
$\sigma$--meson in terms of pion decay constant
$F_\pi$
and the low--energy constants of effective chiral Lagrangian
$L_2$ and $L_3$.
Analysis of these inequlities favours the presence of the light
(mass $ < 750$~ MeV) scalar
%meson
state in the meson spectrum of the multicolour QCD.

As a final remark we note that the sum rules
(\ref{sr}) are derived in the limit of large number of colours,
this implies that the exotic mesons (glueballs, four-quark
states) do not contribute to the sum rules because their
contributions are suppressed by powers of $1/N_c$. This observation
shows that the sum rules (\ref{sr}) can be used for identification
of the nature of various mesons, in particular the low--lying ones.
For example, the sum rules in eq.~(\ref{sr}) tell us that the
leading large--$N_c$ part of the width
(read the width of the $q\bar q$ and hybrid parts)
and the mass of the $\sigma$--meson should satisfy the following
constraint:

\be
\frac{32 \pi F_0^2 M_\sigma^2 \Gamma\left(
\sigma\to\pi\pi \right)}{3 \sqrt{M_\sigma^2-4 m_\pi^2} \left[
M_\sigma^2-2 m_\pi^2
\right]^2} \le 1-\frac{4\left( M_\rho^2-2 m_\pi^2\right)}{F_0^2} L_2 \, .
\ee
Obviously, other sum rules in eq.~(\ref{sr}) and an additional
information about resonance spectrum would provide more
sofisticated constraints on the parameters of $q\bar q$ component
of the $\sigma$--meson. We shall analyse them elsewhere.

\vspace{0.2cm}
\noindent
{\bf Acknowledgements}\\
We are thankful to Shi-Lin~Zhu  for valuable remarks.
This work was supported in part by the Russian Foundation for
Basic Researches (grants RFBR 00-15-96610 and RFBR 01-02-17152),
Russian Ministry of Education (grant E00-3.3-208) and by INTAS
(INTAS Call 2000 project 587).

\end{document}